\documentclass[fleqn,twoside]{article}%
\usepackage{amsmath}
\usepackage{amsfonts}
\usepackage{amssymb}
\usepackage{graphicx}%
\usepackage{caption}
\def\be{\begin{equation}}

\def\ee{\end{equation}}
\def\bes{\begin{equation}\begin{split}&}
\def\es{\end{split}}
\def\bi{\bibitem}

\begin{document}

\title{Conflict between some higher-order curvature invariant terms.}

\author{Dalia saha$^1$,Mohosin Alam$^2$,Ranajit Mandal$^3$,Abhik Kumar Sanyal$^4$}
\maketitle

\noindent

\begin{center}
\noindent
$^{1, 4}$ Dept. of Physics, Jangipur College, Murshidabad, West Bengal, India - 742213\\
\noindent
${^2}$ Dept. of Physics, Saidpur U. N. H. S., Murshidabad, West Bengal, India - 742225. \\
\noindent
${^3}$ Dept. of Physics, Rammohan College, Kolkata, West Bengal, India - 700009.\\

\end{center}

\footnotetext[1]{

\noindent Electronic address:\\
$^1$daliasahamandal1983@gmail.com\\
$^2$alammohosin@gmail.com\\
$^3$ranajitmandalphys@gmail.com\\
$^4$sanyal\_ ak@yahoo.com\\}

\abstract{A viable quantum theory does not allow curvature invariant terms of different higher orders to be accommodated in the gravitational action. We show that there is indeed a conflict between the curvature squared and Gauss-Bonnet squared terms from the point of view of hermiticity. This means one should choose either, in addition to the Einstein-Hilbert term, but never the two together. We explore early cosmic evolution with Gauss-Bonnet squared term.}

\section{Introduction}
The problem associated with bare cosmological constant and the absence of a scalar field in the late universe, motivated cosmologists to propose several curvature induced gravity models, for solving the cosmic puzzle encountered at the late-stage of cosmological evolution. In this context, $F(R, \mathcal{G})$ theory ($R$ and $\mathcal{G}$ are the Ricci scalar and the Gauss-Bonnet term respectively), has been studied largely in recent years, and therefore is one of the prevalent models. It is well-known that the Gauss-Bonnet term is topologically invariant in $4$-dimension. Thus,  contribution from such a term in the field equations requires dilatonic coupling. A dilaton-like scalar field might have existed in the early universe, but no trace of it has been found in the late, low energy regime. On the contrary, if powers of the Gauss-Bonnet term different from one, is taken into account, neither a dilatonic coupling is required nor the pathology of branched Hamiltonian appears \cite{I1,I2,I3,I4,I5,I6,I7}, although in the process, the beauty with second order field equations is sacrificed. In any case, string inspired $F(R,\mathcal{G})$ gravity theory has therefore been regarded as an alternative to the dark energy \cite{I8,I81,I82,I83,I9,I10,I11,I12,I13,I14,I15}, and the viability of $F(R, \mathcal{G})$ model has been examined over years from different angles \cite{I8,I81,I82,I83,I12,I18,I19,I20,I21,I22,I23,I24,I25,I26,I28}. It has been revealed that in the pressureless-dust dominated epoch, a possible transition from deceleration to acceleration, with effective quintessence and even phantom phases can naturally emerge in this framework, without the need to introduce scalar fields. It has also been demonstrated that $F(R,\mathcal{G})$ gravity theory is perfectly viable, since it shows no correction to Newton’s law in flat space for an arbitrary choice of $F(\mathcal{G})$ as well as no instabilities, and thus it is compliant with the stringent solar system constraints \cite{I8}. In particular, the one-loop effective action of $F(\mathcal{G})$ gravity has been found in the de Sitter background, which was then used to derive stability criteria \cite{I83}. In view of all the above cited literatures, it is quite apparent that $F(R, \mathcal{G})$ gravity may be treated as a reasonably good alternative to dark energy. However, the theory has not been contemplated in the context of the evolution of the very early universe, particularly in regard of inflation, which is our current motivation.\\

It was claimed that the form $F(\mathcal{G}) = a_0 \mathcal{G}^n + b_0 \mathcal{G}^m$ can accommodate both early inflation (if $n > 1$) and late time acceleration, (if $m < {1\over 2}$) \cite{I82}. So to study the early universe, it suffices to consider only the first term, with $n > 1$, since the other term contributes insignificantly. Recently, for the sake of simplicity and to get a deeper insight, instead of considering arbitrary power of the Gauss-Bonnet term, an action with Gauss-Bonnet squared term $F(\mathcal{G}) \propto \mathcal{G}^2$ was considered in association with the Einstein-Hilbert action in the following form \cite{I30},

\be\label{A0} A_0 = \int \left[\alpha R + \gamma \mathcal{G}^2\right]\sqrt{-g} d^4 x.\ee
The above action was then probed in the context of early universe, viz. canonical quantization, semiclassical approximation and in the study of inflation. A host of pathologies appeared, some of which were alleviated under the inclusion of a bare cosmological constant term \cite{I30}. However, the problem encountered while studying inflation, particularly the vanishing of the slow-roll parameters, to be precise, could not be mitigated. In order to resolve the pathology, we therefore introduce a scalar field in the present work, that might have existed in the early universe in the form of Higgs boson, which has played a vital role during inflation \cite{I31,I32,I33,I34}. The Higgs boson is measured to have a mass of about $126$ GeV, having spin zero and positive parity. It is well known that the Higgs boson $h$ is an integral part of the `Standard Model' of particle physics and provides a mechanism by which the `Standard Model' particles acquire their mass. The basic idea of Higgs inflation is to identify the Higgs boson $h$ with the cosmic inflaton field $\phi$, thereby establishing a direct connection between elementary particle physics and inflationary cosmology. Hence the action in the presence of a cosmological constant is modified to,

\be\label{A01} A_{1} = \int \left[\alpha(R - 2\Lambda) + \gamma\mathcal{G}^2 - {1\over 2}\phi_{,\mu}\phi^{,\mu} - V(\phi)\right]\sqrt{-g}~d^4x.\ee
However, note that in the above action, Gauss-Bonnet squared term $\mathcal{G}^2 = (R^2 - 4 R_{\alpha\beta}R^{\alpha\beta}+ R_{\alpha\beta\gamma\delta}R_{\alpha\beta\gamma\delta})^2$ contains curvature terms starting from fourth degree. To be more precise, Gauss–Bonnet squared term ($\mathcal{G}^2$) effectively plays the same role as $R^4$ term in the modified theory of gravity, at least in the background of homogeneous and isotropic Robertson–Walker space-time \cite{I12, I30}. Thus the fact that the above action \eqref{A01} skips second degree terms ($R^2, R_{\alpha\beta}R^{\alpha\beta}~ etc.$) and jumps over from the linear sector ($R$) to fourth degree curvature terms ($R^4, R^2 \times R_{\alpha\beta}R^{\alpha\beta}~etc.$), is not seemingly pleasant. For higher powers of Gauss-Bonnet term, situation is even worse. It is important to mention that the effective weak field limits of string, supergravity and other theories lead to an action containing $R^2$ term. Further, $F(R, \mathcal{G})$ model does not exclude curvature squared term. We therefore modify action \eqref{A01} by including a curvature squared term, which now reads as:

\be\label{A} A = \int \left[\alpha(\phi)(R - 2\Lambda) +\beta{R}^2 + \gamma\mathcal{G}^2 - {1\over 2}\phi_{,\mu}\phi^{,\mu} - V(\phi)\right]\sqrt{-g}~d^4x,\ee
where, we have considered functional form of the coupling parameter $\alpha = \alpha(\phi)$ for non-minimal generalization of the action. Above action is under present study for the early universe, inflation in particular. Inflation is essentially a quantum theory of perturbation, which occurred just after the Planck's era between $10^{-42} - 10^{-26} sec$. It is to be mentioned that most of the important physics, close to the Planck's era, may be extracted from the classical action itself, only if the theory admits a reasonably viable quantum dynamics. By viability we mean: a hermitian effective Hamiltonian, if not unitary, and a smooth passage of the quantum theory to the classical domain via a standard semiclassical approximation. It is therefore required to build quantum counterpart of the above action \eqref{A}, at the first place. In the following section, we therefore apply Dirac's constraint algorithm, to make canonical formulation of the action \eqref{A} under consideration, which is a pre-curser to canonical quantization. Next we shall quantize and study the viability of the quantum theory. It will be shown that the hermiticity of the effective Hamiltonian as well as the continuity equation do not allow both the $R^2$ and $\mathcal{G}^2$ terms in the action. In section 3, we therefore drop $R^2$ term (as it has been extensively studied earlier), and study the action in regard of the early universe, inflation, to be precise. In section 4, we shall intuitively discuss the role of the action in the matter dominated era. Section 5 concludes our work.\\

\section{Canonical formulation and quantization:}
In the homogeneous and isotropic Robertson-Walker metric,

\be\label{RW} ds^2 = -N(t)^2 dt^2 + a(t)^2 \left[ {dr^2\over 1-kr^2} + r^2 d\theta^2 + r^2 \sin^2{\theta} d\phi^2\right],\ee
where $N(t)$ is the lapse function, the Ricci scalar and the Gauss-Bonnet term under the choice of the basic variable $h_{ij} = a^2\delta_{ij} = z\delta_{ij}$, (where $h_{ij}$ is the induced three metric) take the form,

\be\begin{split}& R = \frac{6}{N^2}\left(\frac{\ddot a}{a}+\frac{\dot a^2}{a^2}+N^2\frac{k}{a^2}-\frac{\dot N\dot a}{N a}\right) = {6\over N^2}\left[{\ddot z\over 2z} + N^2 {k\over z} - {1\over 2}{\dot N\dot z\over N z}\right]\\&
\mathcal{G} = {24\over N^3 a^3}(N\ddot a - \dot N\dot a)\Big({\dot a^2\over N^2} + k\Big)= {12\over N^2}\left({\ddot z\over z} - {\dot z^2\over 2 z^2} -{\dot N\dot z\over N z}\right)\left({\dot z^2\over 4N^2 z^2} + {k\over z}\right).\end{split}\ee

In the spatially flat space $k = 0$, field equations corresponding to the action (\ref{A}) read as,

\be\label{Mzvariation}\begin{split} \\ & 2\alpha\left({\ddot z\over z}-{\dot z^2\over 4z^2} - \Lambda\right)+2\alpha'\left({\ddot\phi}+{\dot\phi}{\dot z\over z} \right)+2\alpha''{\dot\phi^2}+12\beta\left[\frac{\ddddot z}{z}-\frac{\dddot z \dot z}{z^2}-\frac{3{\ddot z}^2}{4z^2}+\frac{3\ddot z{\dot z}^2}{4z^3}\right]\\&+12\gamma\left[{\dot z^4{\ddddot z}\over z^5}+{8\dot z^3{\ddot z}{\dddot z}\over z^5}-{9\dot z^5{\dddot z}\over z^6}+{6\dot z^2\ddot z^3\over z^5}-{135 \dot z^4\ddot z^2 \over 4 z^6}+{159 \dot z^6\ddot z\over 4z^7} - {195 \dot z^8\over 16 z^8}\right]\\&
= -p -\Big[{1\over 2}\dot\phi^2 - V(\phi)\Big].\end{split}\ee
\be\label{M00}\begin{split}\\& 2\alpha \left({3\dot z^2\over 4z^2}-\Lambda\right)+{3\alpha'\dot\phi \dot z\over z}+18\beta\left(\frac{\dddot z \dot z}{z^2}-\frac{\ddot z \dot z^2}{2z^3}-\frac{\ddot z^2}{2z^2}\right) + 18\gamma\left[{\dot z^5 \dddot z\over z^6} + {3\dot z^4 \ddot z^2\over 2z^6} - {9 \dot z^6\ddot z\over 2 z^7} + {15 \dot z^8\over 8 z^8}\right]\\&= \rho + {1\over 2}\dot\phi^2 + V(\phi).\end{split}\ee
\be\label{Mphivariation} \begin{split} \\& \ddot\phi + {3\over 2}{\dot z\over z}\dot\phi + V'-{3\alpha'\ddot z\over z}+2\Lambda\alpha'= 0.\end{split}\ee
In the above, $p$ is the thermodynamic pressure of a barotropic fluid and $\rho$ is the matter density. It is important to mention that, in contrast to the earlier claim that de Sitter universe naturally occurs at the early or late times in such models \cite{I11}, here we observe that for $\alpha = \alpha_0 =$ constant, the above set of classical field equations does not admit either de-Sitter or power law solutions in the early vacuum dominated ($p = \rho = 0$) era. This is the reason for incorporating $\alpha(\phi)$ instead. The inflationary solution of the above classical field equations, for $\alpha = \alpha(\phi)$ is found in the following standard de-Sitter form,

\be\label{aphi} a=a_0e^{\lambda t}; ~~~ \phi= \phi_0e^{-\lambda t},\ee
As a result, the contribution of terms associated with the coefficient of $\beta$ identically vanishes, and the above field equations \eqref{M00} and \eqref{Mphivariation} reduce to,

\be \label{F1}6\alpha\lambda^2-6\alpha'\phi\lambda^2-576\gamma\lambda^8-2\alpha\Lambda-V-{{\lambda^2\phi^2}\over{ 2}}=0, \ee
\be \label{F2} -2\lambda^2\phi-12\alpha'\lambda^2+2\alpha'\Lambda+V'=0.\ee
Here, $\mathrm{H}\equiv \frac{\dot a}{a}$ denotes the Hubble expansion rate, as usual. From the above set of equations \eqref{F1} and \eqref{F2}, we find the following form of coupling parameters $\alpha(\phi)$, $\gamma(\phi)$ and the potential $V(\phi)$:\\
\be \begin{split}\label{param} & \alpha =-\left[96\gamma\lambda^6+\frac{V_0}{6\lambda^2}+\frac{V_1}{6\lambda^2\phi}+ {\phi^2\over 12}\right]=\left[\alpha_0-\frac{\alpha_1}{\phi}-{\phi^2\over 12}\right],\\& V=\lambda^2\phi^2+12\alpha\lambda^2-2\alpha\Lambda+V_0=\left[-1152\gamma\lambda^8-2{V_1\over \phi}-V_0+2\Lambda\left(96\gamma\lambda^6+\frac{V_0}{6\lambda^2}+\frac{V_1}{6\lambda^2\phi}+ {\phi^2\over 12}\right)\right]\\&=\mathcal{V}_0+{\mathcal{V}_1\over \phi}+\mathcal{V}_2\phi^2,\end{split}\ee
where, the constants are restricted to, $\alpha_0 = -\big(96\gamma\lambda^6+\frac{V_0}{6\lambda^2}\big),~ \alpha_1=\frac{V_1}{6\lambda^2},$ together with $\mathcal{V}_0=-1152\gamma\lambda^8-V_0+2\Lambda\left(96\gamma\lambda^6+\frac{V_0}{6\lambda^2}\right),
~~\mathcal{V}_1=-2{V_1}+\Lambda\frac{V_1}{3\lambda^2},~~\mathcal{V}_2=\frac{\Lambda}{6}.$
Note that, in order to ensure $\alpha > 0$, we must have $\alpha_0 > 0$, i.e. $V_0 < 0$ and $\gamma < 0$. Let us now proceed to find the phase-space structure of the Hamiltonian, corresponding to the action \eqref{A}, in the homogeneous and isotropic minisuperspace \eqref{RW}, which is a precursor towards canonical quantization. The action \eqref{A} reads as,

\be\begin{split} A& = \int \Bigg[ {6\alpha(\phi)\over N^2}\Big({\ddot z\over 2z} + N^2 {k\over z} - {1\over 2}{\dot N\dot z\over N z}\Big)+\frac{9\beta}{N^4}\bigg{(}\frac{\ddot z^2}{z^2}-\frac{2\dot z \ddot z \dot N}{N z^2}+\frac{\dot z^2\dot N^2}{z^2N^2}+\frac{4k{\ddot z}N^2}{z^2}+\frac{4k^2N^4}{z^2}-\frac{4kN\dot N\dot z}{z^2} \bigg{)}\\& +{144\gamma\over N^4}\Big({\ddot z\over z} - {\dot z^2\over 2 z^2} -{\dot N\dot z\over N z}\Big)^2\Big({\dot z^2\over 4N^2 z^2} + {k\over z}\Big)^2 +{1\over 2 N^2}\dot \phi^2 - V(\phi) - 2\Lambda \alpha(\phi)\Bigg]Nz^{3\over 2} dt \int d^3 x,\end{split}\ee
or more explicitly,

\be\begin{split}\label{Act} A = &\int \Bigg[{3\alpha(\phi)\sqrt z\ddot z\over N} + {6\alpha(\phi)\over N}\Big(kN^2 \sqrt{z} - {\dot N\sqrt z\dot z\over 2N}\Big) - 2\alpha(\phi)\Lambda N z^{3\over 2} + N z^{3\over 2}\Big({1\over 2 N^2}\dot\phi^2 - V(\phi)\Big)\\&+\frac{9\beta}{\sqrt z}\bigg{(}\frac{\ddot z^2}{N^3}-\frac{2\dot z \ddot z \dot N}{N^4}+\frac{\dot z^2\dot N^2}{N^5}+\frac{2k{\dot z}^2}{N z}+4k^2N \bigg{)} + 144\gamma\Bigg\{{\ddot z^2\over 16 N^3 z^{9\over 2}}\Big({\dot z^2\over N^2} + 4 k z\Big)^2\\&
-\ddot z\Big({\dot z^6 \over 16 N^7 z^{11\over 2}}+{\dot N\dot z^5\over 8 N^8 z^{9\over 2}}+{k\dot z^4\over 2 N^5z^{9\over 2}}+{k\dot N\dot z^3\over N^6 z^{7\over 2}}+{k^2\dot z^2\over N^3z^{7\over 2}}+{2k^2\dot N\dot z\over N^4 z^{5\over 2}}\Big)
+ {\dot z^8 \over 64 N^7 z^{13\over 2}}\\&+{\dot N\dot z^7\over 16 N^8 z^{11\over 2}} + {\dot N^2\dot z^6\over 16 N^9 z^{9\over 2}}
+{k\dot z^6\over 8 N^5z^{11\over 2}}+{k\dot N\dot z^5\over 2 N^6 z^{9\over 2}} + {k\dot N^2\dot z^4\over 2 N^7 z^{7\over 2}}
+{k^2\dot z^4\over 4N^3z^{9\over 2}}+{k^2\dot N\dot z^3\over N^4 z^{7\over 2}}+{k^2\dot N^2\dot z^2\over N^5 z^{5\over 2}}
\Bigg\}\Bigg]dt.\end{split}\ee
Up on removing divergent terms following integration by parts, the above action \eqref{Act} takes the following form,

\be\begin{split} \label{A2} &A = \int \Bigg[{6\alpha(\phi) N}\Big(-{\dot z^2\over 4 N^2\sqrt z}+ k  \sqrt z - {\Lambda\over 3} z^{3\over 2}\Big)-\frac{3\alpha'(\phi) \dot\phi \dot z\sqrt z}{N}+\frac{9\beta}{\sqrt z}\bigg{(}\frac{\ddot z^2}{N^3}-\frac{2\dot z \ddot z \dot N}{N^4}+\frac{\dot z^2\dot N^2}{N^5}+\frac{2k{\dot z}^2}{N z}\\&+4k^2N \bigg{)}+ N z^{3\over 2}\Big({1\over 2 N^2}\dot\phi^2 - V(\phi)\Big) +144\gamma\Bigg\{{\ddot z^2\over 16 N^3 z^{9\over 2}}\Big({\dot z^2\over N^2} + 4 k z\Big)^2- {15\dot z^8 \over 448 N^7 z^{13\over 2}} + {\dot N^2\dot z^6\over 16 N^9 z^{9\over 2}}
-{13 k\dot z^6\over 40 N^5z^{11\over 2}}\\& + {k\dot N^2\dot z^4\over 2 N^7 z^{7\over 2}}
-{11 k^2\dot z^4\over 12 N^3z^{9\over 2}} +{k^2\dot N^2\dot z^2\over N^5 z^{5\over 2}}
-\ddot z\Big({\dot N\dot z^5\over 8 N^8 z^{9\over 2}}+{k\dot N\dot z^3\over N^6 z^{7\over 2}}+{2k^2\dot N\dot z\over N^4 z^{5\over 2}}\Big)\Bigg\}\Bigg]dt. \end{split}\ee
It is important to notice that unlike `general theory of relativity' the lapse function $N(t)$, which must not have dynamics, and is supposed to act only as a Lagrange multiplier, appears with its time derivative in the above action. However, if a change of variable $\dot z = N x$ is requested, so that a pair of basic variables $h_{ij} = z^2 \delta_{ij}, ~K_{ij} = -{\dot h_{ij}\over 2N} = -{a\dot a\over N}\delta_{ij} = -{\dot z\over 2N}\delta_{ij}$ are addressed, where $K_{ij}$ is the extrinsic curvature tensor, then such incompatibility disappears, as it is manifest in the following action,

\be\begin{split}\label {A3}& A = \int \Bigg[-{6\alpha(\phi) N}\Big({x^2\over 4\sqrt z} - k\sqrt z + {\Lambda\over 3} z^{3\over 2}\Big)-{3\alpha'(\phi) \dot\phi x\sqrt z} +\frac{9\beta}{\sqrt z}\bigg{(}\frac{\dot x^2}{N}+\frac{2kN x^2}{z}+4k^2N \bigg{)}\\&+ N z^{3\over 2}\Big({1\over 2 N^2}\dot\phi^2 - V(\phi)\Big)
+ 144\gamma\Bigg\{{(x^2 + 4 k z)^2\dot x^2\over 16 N z^{9\over 2}} - N\Big({15 x^8 \over 448 z^{13\over 2}} + {13 k x^6\over 40 z^{11\over 2}}+ {11 k^2 x^4\over 12 z^{9\over 2}}\Big)\Bigg\}\Bigg]dt.\end{split}\ee
The reason for introducing none other than the basic variables ($h_{ij}, ~K_{ij}$) is thus established. This is indeed a big triumph of Ostrogradsky's formalism \cite{Ostro1, Ostro2}, since it was only developed to handle higher order theories of classical mechanics long back, but has been found to be well suited for higher-order theory of gravity, as well. Going back, one can observe that neither $\dot z$ nor $\dot N$ appears in the above action, signalling that the corresponding momenta are constrained to vanish. Thus the Hessian determinant also vanishes, implying that the associated point Lagrangian is singular and therefore Ostrogradsky's formalism does not work and Dirac's constrained analysis \cite{Dirac1, Dirac2} is called upon.

\subsection{Analysis of constraint:}

The point Lagrangian may now be expressed in the form,

\be\begin{split}\label {L} &L = -{6\alpha(\phi) N}\Big({x^2\over 4\sqrt z} - k\sqrt z + {\Lambda\over 3} z^{3\over 2}\Big)-{3\alpha'(\phi) \dot\phi x\sqrt z}+\frac{9\beta}{\sqrt z}\bigg{(}\frac{\dot x^2}{N}+\frac{2kN x^2}{z}+4k^2N \bigg{)}
\\&+ N z^{3\over 2}\Big({1\over 2 N^2}\dot\phi^2 - V(\phi)\Big)+ 144\gamma\Bigg\{{(x^2 + 4 k z)^2\dot x^2\over 16 N z^{9\over 2}}- N\Big({15 x^8 \over 448 z^{13\over 2}} + {13 k x^6\over 40 z^{11\over 2}} + {11 k^2 x^4\over 12 z^{9\over 2}}\Big)\Bigg\}+ u\left({\dot z\over N} - x\right),\end{split}\ee
where we have treated the expression $\big{(}{\dot z\over N} - \dot x\big{)}$ as a constraint and incorporated it through the Lagrange multiplier $u$ in the above point Lagrangian. The canonical momenta are,
\be\label{p1}\begin{split} & p_x =\left[{288\gamma\over N}\left({x^4\over 16 z^{9\over 2}}+{k x^2\over 2 z^{7\over 2}}+{k^2\over z^{5\over 2}}\right) +{18\beta\over N\sqrt z}\right]\dot x,\hspace{0.3 in} p_z = {u\over N},\\& p_{\phi}=-{3\alpha'x\sqrt z}+{z^{3\over 2}\dot\phi\over N},\hspace{0.3 in}p_N = 0 = p_u. \end{split}\ee
Therefore the primary Hamiltonian reads as,

\be\label{Hp1}\begin{split} H_{p1}= &N\Bigg{[}{p_x^2\over {576\gamma \left({x^4\over 16 z^{9\over 2}}+{k x^2\over 2 z^{7\over 2}}+{k^2\over z^{5\over 2}}\right)+\frac{36\beta}{\sqrt z}}}+ {6\alpha}\Big({x^2\over 4\sqrt z} - k\sqrt z + {\Lambda\over 3} z^{3\over 2}\Big)\\&+36\gamma x^4 \left({15x^4\over 112 z^{13\over 2}}+{13k x^2\over 10z^{11\over 2}}+{11k^2\over 3 z^{9\over 2}}\right)
-\frac{18k\beta}{\sqrt z}\bigg{(}\frac{x^2}{z}+2k\bigg{)}+{p_{\phi}^2\over 2z^{3\over 2}}\\&+{3\alpha' x p_{\phi}\over z}+{9\alpha'^2 x^2\over 2\sqrt z}+Vz^{3\over 2}\Bigg{]}+u_1\big{(}Np_z-u\big{)}+u_2 p_u+ux.\end{split}\ee
The definition of momenta \eqref{p1} reveals that we need to consider two primary constraints involving Lagrange multipliers or their conjugates viz,

\be\label{constraints}\phi_1 = Np_z-u \approx 0,~ \phi_2 = p_u \approx 0.\ee
Note that the constraint $\phi_3 = p_N$ associated with lapse function $N$ vanishes strongly, since it is non-dynamical, and therefore has been safely ignored. The above two primary constraints \eqref{constraints} are second class, since, they have non-vanishing Poisson bracket with  other constraints $\{\phi_1, \phi_2\} \ne 0$. In two possible ways the second-class constraints may be handled. First, the Hamiltonian may be extended by adding the constraints with arbitrary Lagrange multipliers, which may be determined  unambiguously, solving the consistency equations due to the fact that $|{\phi_i,\phi_j}| \neq 0$. Second, Dirac bracket may be introduced and the constraints are thrown away. Since, appropriate commutation relations during transition to the quantum theory follow from Dirac brackets, therefore here we compute Dirac brackets first and then follow the first method, which is straight forward. The Dirac bracket of two functions $f$ and $g$ in phase space is defined as.

\be \big\{f,g\big\}_{DB} = \big\{f,g\big\}_{PB} - \sum_{ij}\big\{f,\phi_i\big\}_{PB}M^{-1}_{ij}\big\{\phi_j,g\big\}_{PB},\ee
where the matrix $M_{ij} = \big\{\phi_i,\phi_j\big\}_{PB}$, has its inverse denoted by $M^{-1}_{ij}$. In the present case, the matrix and its inverse are simply
\be M_{ij} = \left(
            \begin{array}{cc}
              0 & -1 \\
              1 & 0 \\

            \end{array}
          \right)
~~ \mathrm{and}~~ M^{-1}_{ij} = \left(
            \begin{array}{cc}
              0 & 1 \\
              -1 & 0 \\

            \end{array}
          \right)\ee
Therefore, the Dirac bracket reduces to the following form:

\be \big\{f,g\big\}_{DB} = \big\{f,g\big\}_{PB} + \sum_{ij}\epsilon_{ij}\big\{f,\phi_i\big\}_{PB}\big\{\phi_j,g\big\}_{PB},\ee
where $\epsilon_{ij}$ is the Levi-Civita symbol. A straightforward calculation then reveals:

\be\begin{split} \{z,p_z\}_{DB} &= \{z,p_z\}_{PB} + \epsilon_{11}\{z,\phi_1\}_{PB}\{\phi_1,p_z\}_{PB} + \epsilon_{12}\{z,\phi_1\}_{PB}\{\phi_2,p_z\}_{PB}\\&\hspace{2cm}+ \epsilon_{21}\{z,\phi_2\}_{PB}\{\phi_1,p_z\}_{PB} + \epsilon_{22}\{z,\phi_2\}_{PB}\{\phi_2,p_z\}_{PB}\\&
      = \{z,p_z\}_{PB} = 1,\end{split}\ee
 since, $\{\phi_i, p_z\}_{PB} = 0$. Likewise, $\{x,p_x\}_{DB} = \{x,p_x\}_{PB} = 1$,  $\{z,p_x\}_{DB} = \{z,p_x\}_{PB} = 0$,$\{p_z,p_x\}_{DB} = \{p_z,p_x\}_{PB} = 0$. Therefore, the correct implementation of canonical quantization are the standard commutation relations, $[\hat{z}, \hat{p}_z] = i\hbar$, $[\hat{z}, \hat{p}_x] = 0$. The reason behind such equality lies in the fact that, $\phi_2$ strongly vanish. So, while following this prescription, one can throw them away, but not, if one follows the first prescription, since in that case it will not be possible to compute the Lagrange multipliers. Let us now find the phase-space structure of the Hamiltonian following the standard formulation, as mentioned, introducing the constraints $\phi_1=Np_z-u\approx 0;$ and $\phi_2=p_u\approx 0$ through the Lagrange multipliers $u_1$ and $u_2$ respectively, so that the modified primary Hamiltonian takes the following form,

\be\begin{split}  H_{p2}= H_{p1}+u_1\big{(}Np_z-u\big{)}+u_2 p_u.\end{split}\ee
Note that the Poisson brackets $\{x,p_x\}=\{z,p_z\}=\{\phi,p_{\phi}\}=\{u,p_u\}=1,$ hold. The fact that the constraints should remain preserved in time, is exhibited through the following Poisson brackets,
\be\label{phi12}\begin{split} &\dot\phi_1=\{\phi_1,H_{p1}\}\approx 0\Rightarrow u_2=-N{\partial H_{p1}\over \partial z};\;\; \\&\dot\phi_2=\{\phi_2,H_{p1}\}\approx 0\Rightarrow u_1=x. \end{split}\ee
Therefore the primary Hamiltonian is modified to,
\be\label{HP2} H_{p2} = H_p - Np_u{\partial H_{p1}\over \partial z}.\ee
As the constraint should remain preserved in time in the sense of Dirac, so
\be\label{Dphi2}\begin{split}\dot\phi_2=&\{\phi_2,H_{p2}\}\approx 0, \Rightarrow p_u=0. \end{split}\ee
Thus, finally the phase-space structure of the Hamiltonian, being free from constraints reads as,
\be\label{Hp1c}\begin{split}  & H=N\Bigg{[}x p_z+{p_x^2\over {576\gamma \left({x^4\over 16 z^{9\over 2}}+{k x^2\over 2 z^{7\over 2}}+{k^2\over z^{5\over 2}}\right)+\frac{36\beta}{\sqrt z}}}+{p_{\phi}^2\over 2z^{3\over 2}}+{3\alpha' x p_{\phi}\over z}-\frac{18k\beta}{\sqrt z}\bigg{(}\frac{x^2}{z}+2k\bigg{)}\\&+36\gamma x^4 \left({15x^4\over 112 z^{13\over 2}}+{13k x^2\over 10z^{11\over 2}}+{11k^2\over 3z^{9\over 2}}\right)+{9\alpha'^2 x^2\over 2\sqrt z}+ {6\alpha}\Big({x^2\over 4\sqrt z} - k\sqrt z + {\Lambda\over 3} z^{3\over 2}\Big)+Vz^{3\over 2}\Bigg{]}=N\mathcal{H},\end{split}\ee
and diffeomorphic invariance $H = N\mathcal{H}$ is established. The action (\ref{A3}) may now be expressed in canonical ADM form ($k = 0$) as,
\be \label{ADMH}\begin{split}& A=\int\Big{(}\dot z p_z+\dot x p_x+\dot\phi p_{\phi}-N\mathcal{H}\Big{)}dt d^3x=\int\Big{(}\dot h_{ij}\pi^{ij}+\dot K_{ij}{\Pi}^{ij}+\dot\phi p_{\phi}-N\mathcal{H}\Big{)}dt d^3x,\end{split}\ee
where $\pi^{ij}$ and $\Pi^{ij}$ are momenta canonically conjugate to $h_{ij}$ and $K_{ij}$ respectively. Clearly, the canonical (ADM) formulation is apparent in terms of the basic variables only, and hence the very importance of using basic variables has once again been established. The presence of the $xp_z$ term in the Hamiltonian (\ref{Hp1c}) reveals the fact that as in the case of different higher order theories studied earlier, the present theory also leads to the schrodinger-like equation leading to a quantum mechanical probabilistic interpretation. Since the momentum $p_z$ appeared as a constraint \eqref{p1} in the action we started with, it is required to find its expression for future consideration. This may now be found from the Hamilton's equation. To avoid complications, we express the Hamilton's equation (\ref{Hp1c}) for $k=0$ as,

\be\label{Hp2c}\begin{split} \mathcal{H} = &\Bigg{[}x p_z+{p_x^2\over{36({\gamma x^4\over z^{9\over 2}}+{\beta\over{\sqrt z}})}}+{p_{\phi}^2\over 2z^{3\over 2}}+{3\alpha' x p_{\phi}\over z}+ {6\alpha}\Big({x^2\over 4\sqrt z} + {\Lambda\over 3} z^{3\over 2}\Big)+36\gamma x^4 \left({15x^4\over 112 z^{13\over 2}}\right)\\&+{9\alpha'^2 x^2\over 2\sqrt z}+Vz^{3\over 2}\Bigg{]}.\end{split}\ee

Now from Hamilton's equation, we get

\be\label{px1}\begin{split}&\dot x=\frac{p_x}{18({\gamma x^4 \over {z}^{9\over2}}+{\beta \over \sqrt z})},~~~~\dot z=x,~~~~\dot\phi=\frac{p_{\phi}}{z^{3\over 2}}+\frac{3\alpha'x}{z},\\&
\dot p_x=\bigg{(}-p_z+\frac{\gamma x^3 p_x^2 }{9 {z}^{9\over2}({\gamma x^4\over {z}^{9\over2}}+{\beta\over \sqrt z})^2}-\frac{3\alpha x}{\sqrt z}-\frac{270\gamma x^7}{7z^{13\over2}}-\frac{3\alpha'p_\phi}{z}+\frac{9\alpha'^2x}{\sqrt z}\bigg{)},\\&
 \\&\dot p_z=\Bigg(\frac{p_x^2({9\gamma x^4 z^{-{11\over 2}}}+{\beta z^{-{3\over2}}})}{72({\gamma x^4 z^{-{9\over2}}}+{\beta z^{-{1\over2}}})^2}+\frac{3p_{\phi}^2}{4z^{5\over 2}}+\frac{3 p_{\phi}\alpha'x }{z^2}+\frac{3\alpha x^2}{4z^{3\over 2}}-3\alpha\Lambda z^{1\over 2}+\frac{1755\gamma x^8}{56z^{15\over 2}}\\&
~~~~~~~+\frac{9\alpha'^2 x^2}{4z^{3\over 2}}+\frac{2025x^{14}}{196z^{27\over 2}}-\frac{3V z^{1\over 2}}{2}\Bigg),~~~~~~~~~\dot p_{\phi}=0.\end{split}\ee
From the first relation of the above set of Hamilton's equations, we obtain,

\be\label{px2} p_x=18\left(\frac{\gamma\dot z^4\ddot z}{z^{9\over2}}+{\beta\ddot z\over \sqrt z}\right);~~ \mathrm{and~hence}~~\dot p_x=\frac{18\gamma\dot z^4\dddot z}{z^{9\over2}}+\frac{72\gamma\dot z^3\ddot z^2}{z^{9\over2}}-\frac{81\gamma\dot z^3\ddot z}{z^{11\over2}}+18\left(\frac{\beta{\dddot z}}{\sqrt z}-\frac{\beta \ddot z\dot z}{2z^{3\over 2}}\right).\ee
Equating $\dot p_x$ from the above two equations \eqref{px1} and \eqref{px2}, we find the expression for $p_z$ as,
\be\label{pz}\begin{split}& p_z = -18\gamma\bigg{(}\frac{2\dot z^3\ddot z^2}{z^{9\over2}}+\frac{15\dot z^7}{7z^{13\over2}}+\frac{\dot z^4\dddot z}{z^{9\over2}}-\frac{9\dot z^5\ddot z}{2z^{11\over2}}\bigg{)}-3\alpha'\dot\phi\sqrt z-\frac{3\alpha\dot z}{\sqrt z}+18\left(\frac{\beta \ddot z\dot z}{2z^{3\over 2}}-\frac{\beta{\dddot z}}{\sqrt z}\right).\end{split}\ee
Before ending this subsection, we would like to mention that the modified theory of gravity under consideration is essentially a higher order theory of gravity and as such suffer from Ostrogradsky's instability. Such instability appears due to degeneracy, which may be taken care of by discarding the divergent terms appearing in the action, which we did. For the purpose of canonical formulation, we seek additional degrees of freedom, which is the basic variable ($K_{ij}$). As a result, the action has been written in canonical form and the field equations reduce to second order, with additional equation. In the process, Ostrogradsky's instability disappears.

\subsection{Canonical quantization:}

Canonical quantization of the Hamiltonian \eqref{Hp1c} is now straight forward,
\be\label{Q}\begin{split} &{i\hbar z^{-{9\over 2}}}\frac{\partial\Psi}{\partial z}=-\frac{\hbar^2}{36x[\gamma{(}x^2+4kz{)}^2+\beta z^4]} \bigg{(}\frac{\partial^2}{\partial x^2} +\frac{n}{x}\frac{\partial}{\partial x}\bigg{)}\Psi-\frac{\hbar^2}{2xz^6}\frac{\partial^2\Psi}{\partial \phi^2}+{3 \widehat{\alpha'p_{\phi}}\over z^{11\over 2}}\Psi\\&+\Big{[}{3\alpha(x^2-4kz)\over 2xz^5}+36\gamma x^3 \left({15x^4\over 112 z^{11}}+{13k x^2\over 10z^{10}}+{11k^2\over 3 z^4}\right)+{9\alpha'^2 x\over 2z^5} +{2\alpha \Lambda\over xz^3}-\frac{18k\beta}{z^5}\left({x\over z} +\frac{2k}{x}\right) +{V\over xz^{3}}\Big{]}\Psi.\end{split}\ee
where, $n$ is the operator ordering index. Still there remains some operator ordering ambiguity, particularly between $\alpha'$ and $p_{\phi}$, which may be resolved only after having the specific knowledge regarding the forms of $\alpha(\phi)$. This is available from the classical de-Sitter solutions \eqref{param}, using which we can express equation \eqref{Q} as,

\be \label{qh4C} \begin{split} {i\hbar z^{-{9\over 2}}}\frac{\partial\Psi}{\partial z}&=-\frac{\hbar^2}{36[\gamma x^5+\beta x z^4]}\bigg{(}\frac{\partial^2}{\partial x^2} +\frac{n}{x}\frac{\partial}{\partial x}\bigg{)}\Psi-\frac{\hbar^2}{2xz^6}\frac{\partial^2\Psi}{\partial \phi^2}+\frac{3i\hbar \alpha_1}{z^{\frac{11}{2}}}\bigg{(}\frac{\Psi}{\phi^3}-\frac{1}{\phi^2}\frac{\partial\Psi}{\partial \phi}\bigg{)} +\frac{3i\hbar }{12z^{\frac{11}{2}}} \bigg{(}2\phi\frac{\partial\Psi}{\partial\phi}+\Psi\bigg{)}\\&
    +\bigg{[}\frac{9x}{2z^5} \bigg{(} \frac{\alpha_1^2}{\phi^4}-\frac{\alpha_1}{3\phi}+{\phi^2\over 36 }\bigg{)} -\frac{3 x}{2z^5}\big{(}\alpha_0-\frac{\alpha_1}{\phi}-{\phi^2\over 12}\big{)}+\frac{135\gamma x^7}{28z^{11}} +\frac{1}{xz^3}\big{(}-1152\gamma\lambda^8-2{V_2\over \phi}-V_1\big{)}\bigg{]}\Psi, \end{split}\ee
where Weyl symmetric ordering has been performed carefully between $\alpha'$ and $p_{\phi}$, and $k = 0$ has been set. Now, under a change of variable, the above modified Wheeler-de-Witt equation, takes the look of Schr\"{o}dinger equation, viz.,\\
\be \label{qh4c} \begin{split} {i\hbar}\frac{\partial\Psi}{\partial \sigma}&=-\frac{\hbar^2}{198\left[\gamma x^5+\beta x \sigma^{8\over 11}\right]}\bigg{(}\frac{\partial^2}{\partial x^2} +\frac{n}{x}\frac{\partial}{\partial x}\bigg{)}\Psi-\frac{\hbar^2}{11x\sigma^{\frac{12}{11}}}\frac{\partial^2\Psi}{\partial \phi^2}+\frac{6i\hbar} {11\sigma}\bigg{(}{\phi \over 6}-\frac{\alpha_1}{\phi^2} \bigg{)} \frac{\partial\Psi}{\partial \phi} \\&+\frac{6i\hbar}{11\sigma} \bigg{(}{1\over 12}+{\alpha_1\over \phi^3}\bigg{)}\Psi+V_{e}\Psi=\widehat H_e \Psi, \end{split}\ee
where, $\sigma=z^{\frac{11}{2}}=a^{11}$ plays the role of internal time parameter. In the above equation, the effective potential $V_{e}$ is given by,\\
\be\begin{split} V_{e}=&\bigg{[}\frac{9x}{11\sigma^{\frac{10}{11}}} \bigg{(} \frac{\alpha_1^2}{\phi^4}-\frac{\alpha_1}{3\phi}+{\phi^2\over 36 }\bigg{)} -\frac{3 x}{11\sigma^{\frac{10}{11}}}\big{(}\alpha_0-\frac{\alpha_1}{\phi}-{\phi^2\over 12}\big{)}+\frac{135\gamma x^7}{154\sigma^2} \\&\hspace{3.0 in}+\frac{2}{11x\sigma^{\frac{6}{11}}}\big{(}-1152\gamma\lambda^8-2{V_2\over \phi}-V_1\big{)}\bigg{]}.\end{split}\ee

\subsection{Hermiticity of $\widehat{H}_e$ and probabilistic interpretation:}

We now proceed to establish hermiticity of the Hamiltonian operator $\widehat H_e$, which is essentially the necessary (although not sufficient) requirement for unitary time evolution of quantum dynamics. The effective Hamiltonian $\widehat H_e$ is split for $k=0$ as, $\widehat H_e = \widehat H_1+\widehat H_2+\widehat H_3+\widehat V_{e}$, where,
\be\begin{split}
\widehat H_1 =& -\frac{\hbar^2}{198\Big[\gamma x^5 +\beta x\sigma^{8\over 11}\Big]}\Big{(}\frac{\partial^2}{\partial x^2} +\frac{n}{x}\frac{\partial}{\partial x}\Big{)},\hspace{0.05 in}
\widehat H_2 = -\frac{\hbar^2}{11x\sigma^{\frac{12}{11}}}\Big(\frac{\partial^2}{\partial \phi^2}\Big),\\& \widehat H_3 =  \frac{6i\hbar} {11\sigma}\bigg{(}{\phi \over 6}-\frac{\alpha_1}{\phi^2} \bigg{)} \frac{\partial}{\partial \phi} +\frac{6i\hbar}{11\sigma} \bigg{(}{1\over 12}+{\alpha_1\over \phi^3}\bigg{)},\hspace{0.05 in} \widehat V_{e} = V_{e}.\end{split}\ee
It suffices to establish hermiticity of the Hamiltonian operator $\widehat H_1$, since $\widehat H_2$, $\widehat H_3$ and $\widehat V_{e}$ are trivially hermitian.
\be\label{psi}\begin{split}& \int \big{(}\widehat H_1\Psi\big{)}^*\Psi dx= -\int\frac{\hbar^2}{198\left[\gamma x^5 +\beta x\sigma^{8\over 11}\right]} \bigg{(}\frac{\partial^2\Psi^*}{\partial x^2} +\frac{n}{x}\frac{\partial\Psi^*}{\partial x}\bigg{)}\Psi dx.\end{split}\ee
Under integration by parts twice and dropping the first term due to fall-of condition, we obtain,
\be\label{H1}\begin{split} \int \Big{(}\widehat H_1\Psi\Big{)}^*\Psi dx = &-\frac{\hbar^2}{198} \int \Psi^*\left[\frac{1}{\Big[\gamma{x^5} +\beta x\sigma^{8\over 11}\Big]}\left(\frac{\partial^2\Psi}{\partial x^2}\right) - \frac{{(n+10)\gamma x^4} +(n+2)\beta \sigma^{8\over 11}}{\Big[ \gamma{x^5} +\beta x\sigma^{8\over 11} \Big]^2} \left(\frac{\partial \Psi}{\partial x}\right)\right] \\&+ \frac{\hbar^2}{198} \int\Psi^*\Psi\frac{\partial }{\partial x}\left(\frac{\gamma{x^4}\left(n+5\right) +\beta \sigma^{8\over 11}\left(n+1\right)}{\Big[\gamma{x^5} +\beta x\sigma^{8\over 11}\Big]^2 } \right) dx. \end{split}\ee
In order to proceed further, one should note that equation \eqref{psi} does not contain any term in the form $\Psi^* \Psi$, and therefore to ensure $\widehat H_1$ to be hermitian, primarily one has to get rid of the last term appearing in equation (\ref{H1}). This usually fixes the operator ordering index. However, here we have two options at hand: either, $\gamma = 0$, with, $n = -1$, which eliminates Gauss-Bonnet squared term, or, $n=-5$ with $\beta=0$, which eliminates $R^2$ term. This is our main result: Gauss-Bonnet squared term cannot be coupled with the scalar curvature squared term, from the very fundamental requirement that the Hamiltonian operator has to be hermitian. \\

Now, let us try to establish the continuity equation despite the fact that the Hamiltonian operator \eqref{qh4c} is time-dependent. Defining the probability density $\rho = \Psi^*\Psi$ as usual, we find,
\begin{center}
    \be\begin{split} \label{cont}\frac{\partial\rho}{\partial\sigma}=-&\frac{\partial}{\partial x}\Big{[}\frac{i\hbar}{198\big[\gamma x^5+\beta x\sigma^{8\over 11}\big]}\big{(}\Psi\Psi^*_{,x}-\Psi^*\Psi_{,x} \big{)}\Big{]}-\frac{\partial}{\partial\phi}\Big{[}\frac{i\hbar}{11x\sigma^{\frac{12}{11}}} \big{(}\Psi \Psi^*_{,\phi}-\Psi^*\Psi_{,\phi} \big{)}-\frac{6}{11\sigma}\big{(}{\phi \over 6}-\frac{\alpha_1}{\phi^2} \big{)}\Psi^*\Psi\Big{]}\\&\hspace{1.5 in}-{i\hbar\over 198}\times\frac{ {\big{(}\Psi\Psi^*_{,x}-\Psi^*\Psi_{,x} \big{)}}\big[\gamma{x^4 }\left(n+5\right) +\beta \sigma^{8\over 11}(n+1)\big] }{\big[\gamma x^5+\beta x\sigma^{8\over 11}\big]^2}. \end{split} \ee
\end{center}
Clearly, continuity equation cannot be established unless the last term vanishes. This again requires either the choice $\gamma = 0$, with, $n = -1$, which eliminates Gauss-Bonnet squared term, or, $n=-5$ with $\beta=0$, which eliminates $R^2$ term. Thus, both the curvature squared term and the Gauss-Bonnet squared term simultaneously cannot be present in the action. \\

It has been shown earlier that Gauss–Bonnet squared term ($\mathcal{G}^2$) effectively plays the same role as $R^4$ term in the modified theory of gravity, at least in the background of homogeneous and isotropic Robertson–Walker space-time \cite{I12, I30}. Thus, there is indeed a conflict between the two curvature invariant terms of different orders. Hence we conclude that, from the point of view of hermiticity, and for the purpose of establishing continuity equation (which leads to standard probabilistic interpretation), a gravitational action is forbidden from associating curvature scalars of different (higher) orders. To be more precise, we find a selection rule which states that in a gravitational action, Einstein-Hilbert (E-H) term may be associated only with one higher order term, but not a combination of different orders like $R^2$, $R^3$, $R^4$. Question inevitably arises: how far it is justified to trade hermiticity of the Hamiltonian operator as a fundamental physical requirement. The reason for raising this issue is: if the theory contains ghosts, as shown by Stell ($R_{\mu\nu}$ corresponds to massive spin $2$ field, with negative norms, which are ghosts) \cite{38} and Chiba (any generalized gravity of containing $F(R, R_{ab}R^{ab}, R_{abcd}R^{abcd})$ terms, corresponds to massive spin $2$ field which are ghosts) \cite{39}, then the entire quantum framework appears to run from serious problem, because the evolution in that case is not unitary. We remind that a self adjoint operator is necessarily hermitian, but the reverse is not true. If an operator is self-adjoint then only the dynamics is unitary and the time parameter can be extended to the real line. However, the computations, which made people believe that higher order theories suffer from instabilities and violation of unitarity, are entirely based on perturbative analysis about Minkowski spacetime. There are counter arguments with comprehensive evidence that perturbative analysis might be naive and misleading and full nonperturbative theory will be free from these difficulties \cite{40,41,42,43,44,45,46,47}. Further, although Chiba stated that such theories cannot lead to viable gravity theories from both phenomenological (solar system experiments) and theoretical (consistency) points of view, a class of $F(R, \mathcal{G})$ gravity theory has been shown to pass the solar tests \cite{I82, I9,I10,I11,I15, I19} and also consistent from the point of view of stability \cite{I8,I83,I12,I14,I18,I20,I23,I24,I25}.\\

In fact, the issue of ghosts stems from Ostrogradski's instability, which is typically recognized as breakdown of unitarity by high energy physicists. However, there exists at least three different ways to avoid the Ostrogradskian instability, by violating the assumption of nondegeneracy upon which it is based. These are through partial integration, through gauge invariance, and by imposing constraints so as to make the theory agree with its perturbative development \cite{48}. In this context, note that we have followed the first technique and removed divergent terms from the action under integration by parts. Further, the problem also stems from the fact that the gravitational action corresponding to the General theory of relativity is unbounded from below. Note that, the present analysis is independent of the presence of E-H term. The rest of the action contains $R^2$ and $\mathcal{G}^2$ terms and hence is positive definite like the one chosen earlier by Horowitz \cite{49}. In the process, we believe that the problem of stability as well as unitarity disappear, although there exists different school of thoughts with counter arguments. We therefore leave this debatable issue here.\\

In view of the above discussions, we trade the issue of Hermiticity and continuity equation as fundamental physical requirement, and use these issues as a selection rule for choosing an action. The conclusion is therefore: the fundamental requirements of hermiticity and continuity equation do not allow terms of different higher orders to incorporate into the action. Since we have already handled curvature square term, in the rest of the present work, we therefore concentrate on the viability of Gauss-Bonnet squared term in the context of early universe. For this reason, let us disregard $R^2$ term from action \eqref{A}, choosing $\beta=0$, and proceed to test the viability of the following action,

\be\label{AG} A_1 = \int \left[\alpha(\phi)(R - 2\Lambda) + \gamma\mathcal{G}^2 - {1\over 2}\phi_{,\mu}\phi^{,\mu} - V(\phi)\right]\sqrt{-g}~d^4x,\ee
in connection with inflation and its evolution in the matter dominated eras. It is important to mention that the same classical de-Sitter solutions \eqref{aphi} under the conditions \eqref{param} hold, since $R^2$ term identically vanishes under such choice. The Hamiltonian \eqref{Hp1c} therefore takes the form,

\be\label{woR2}\begin{split}\mathcal{H}= x p_z+&{p_x^2\over {576\gamma \left({x^4\over 16 z^{9\over 2}}+{k x^2\over 2 z^{7\over 2}}+{k^2\over z^{5\over 2}}\right)}}+{p_{\phi}^2\over 2z^{3\over 2}}+{3\alpha'xp_{\phi}\over z}+36\gamma x^4 \left({15x^4\over 112 z^{13\over 2}}+{13k x^2\over 10z^{11\over 2}}+{11k^2\over 3z^{9\over 2}}\right)\\&+ {6\alpha}\Big({x^2\over 4\sqrt z} - k\sqrt z + {\Lambda\over 3} z^{3\over 2}\Big)+{9\alpha'^2x^2\over 2\sqrt z}+Vz^{3\over 2},\end{split}\ee
and the corresponding modified Wheeler-DeWitt equation is,

\be \label{Mqn} \begin{split} &{i\hbar z^{-{9\over 2}}}\frac{\partial\Psi}{\partial z}=-\frac{\hbar^2}{36x[\gamma{(}x^2+4kz{)}^2]} \bigg{(}\frac{\partial^2}{\partial x^2} +\frac{n}{x}\frac{\partial}{\partial x}\bigg{)}\Psi-\frac{\hbar^2}{2xz^6}\frac{\partial^2\Psi}{\partial \phi^2}+{3 \widehat{\alpha'p_{\phi}}\over z^{11\over 2}}\Psi\\&+\Big{[}{3\alpha(x^2-4kz)\over 2xz^5}+36\gamma x^3 \left({15x^4\over 112 z^{11}}+{13k x^2\over 10z^{10}}+{11k^2\over 3 z^4}\right)+{9\alpha'^2 x\over 2z^5} +{2\alpha \Lambda\over xz^3} +{V\over xz^{3}}\Big{]}\Psi. \end{split}\ee
Equation \eqref{Mqn} may be cast in the following Schr\"odinger like equation, using classical solution \eqref{param} in flat space ($k = 0$) and applying Weyl operator ordering, as before,

\be \label{qh4cR} \begin{split} {i\hbar}\frac{\partial\Psi}{\partial \sigma}&=-\frac{\hbar^2}{198\gamma x^5}\bigg{(}\frac{\partial^2}{\partial x^2} +\frac{n}{x}\frac{\partial}{\partial x}\bigg{)}\Psi-\frac{\hbar^2}{11x\sigma^{\frac{12}{11}}}\frac{\partial^2\Psi}{\partial \phi^2}+{6i\hbar\over 11\sigma}\bigg( {\phi\over 6}-{\alpha_1\over \phi^2} \bigg){\partial \Psi\over \partial\phi}\\& +{6i\hbar\over 11\sigma}\bigg( {1\over 12}+{\alpha_1\over \phi^3}\bigg)\Psi +V_{e}\Psi=\widehat H_e \Psi. \end{split}\ee
Further, if we choose, $n=-5$, then in view of (\ref{H1}) we find,

\be\label{H1.1}\begin{split} &\int \big{(}\widehat H_1\Psi\big{)}^*\Psi dx=-\frac{\hbar^2}{198\gamma}\int \Psi^*\bigg{[}\frac{1}{x^5}\frac{\partial^2\Psi}{\partial x^2} -\frac{5}{x^6}\frac{\partial\Psi}{\partial x}\bigg{]}dx=\int \Psi^*\widehat H_1\Psi dx. \end{split}\ee
Thus $\widehat H_1$ is hermitian, and so is the effective Hamiltonian operator $\widehat H_e$. The continuity equation \eqref{cont} can now be expressed as,

\be \begin{split}&\frac{\partial\rho}{\partial\sigma}+\frac{\partial{J}_x}{\partial x}+\frac{\partial{J}_z}{\partial z}+\frac{\partial{J}_\phi}{\partial \phi} = \frac{\partial\rho}{\partial\sigma}+ \nabla . \mathbf{J}=0, \\&
{J}_x = \frac{i\hbar}{198\gamma x^5}\big{(}\Psi\Psi^*_{,x}-\Psi^*\Psi_{,x} \big{)},
\hspace{.1 in} J_z=0, \hspace{0.1 in} {J}_{\phi} =  \frac{i\hbar}{11x\sigma^{\frac{12}{11}}} \big{(}\Psi \Psi^*_{,\phi}-\Psi^*\Psi_{,\phi} \big{)}-{6\over 11\sigma}\bigg({\phi\over 6}-{\alpha_1\over \phi^2}\bigg)\Psi^*\Psi.\end{split}\ee
where, $\mathbf{J}=(J_x, J_z,J_{\phi} )$ is current density. Thus, conservation of probability is ensured.\\

The quantum description of the universe must finally lead to an appropriate semiclassical wavefunction under suitable semiclassical (WKB) approximation, so that classical universe we live in, emerges. In this context let us recall Hartle criterion for the selection of classical trajectories \cite{Hartle}. It states that: if the approximate wavefunction obtained following some appropriate semiclassical approximation is strongly peaked, then there exists correlations among the geometrical and matter degrees of freedom, and the emergence of classical trajectories (i.e. the universe) is expected, on the contrary, if it is not peaked, correlations are lost'. In the present case, the semiclassical wavefunction is found as [see appendix for detailed computation],

\be\label{Psi4}\Psi = \Psi_0e^{\frac{i}{\hbar}S(x,z,\phi)}=\Psi_{01}e^{{i\over \hbar}\Big[-4\alpha_0\lambda z^{3\over 2}+{6\alpha_1 \lambda z^2\over a_0\phi_0}+{3840\gamma \lambda^7 z^{3\over 2}\over 7}\Big]},\ee
where, \be \Psi_{01}=\Psi_0e^{-F(z)}.\ee
Thus, the wave function is oscillatory, which implies that the semiclassical wavefunction is strongly peaked around the classical inflationary solution and therefore according to Hartle prescription \cite{Hartle}, emergence of classical trajectory is confirmed. Since the present quantum theory admits a viable semiclassical approximation, most of the important physics may be extracted from the classical action itself, as already mentioned. We therefore study inflation in view of the classical field equations, next.

\subsection{Inflation under slow roll approximation:}

Instead of standard slow roll parameters, here we use a combined hierarchy of Hubble and coupling flow parameters \cite{In1, In2, In3, In4, In5, In6, In7, In8} as follows. Firstly, we describe the background evolution by a set of horizon flow functions (the behaviour of Hubble distance during inflation) starting from,

\be \label{dh}\epsilon_0=\frac{d_{\mathrm{H}}}{d_{\mathrm{H}_i}},\ee
where $d_{\mathrm{H}}=\mathrm{H}^{-1}$ is the Hubble distance, also called the horizon in our chosen units. We use suffix $i$ to denote the era at which inflation was initiated. Now hierarchy of functions is defined in a systematic way as,

\begin{center}
    \be \label{el} \epsilon_{l+1}=\frac{d\ln|\epsilon_l|}{d\mathcal{N}},~~l\geq 0. \ee
\end{center}
In view of the definition of the number of e-fold expansion, $\mathcal{N}=\ln{\big(\frac{a}{a_i}\big)}$, which implies $\dot {\mathcal{N}}=\mathrm{H},$ we compute $\epsilon_1=\frac{d\ln{d_{\mathrm{H}}}}{d\mathcal{N}},$ which is the logarithmic change of Hubble distance per e-fold expansion $\mathcal{N}$, and is known as the first slow-roll parameter: $\epsilon_1=\dot{d_{\mathrm{H}}}=-\frac{\dot {\mathrm{H} }}{\mathrm{H}^2}$. This implies that the Hubble parameter $\mathrm{H}$ almost remains constant during inflation. In view of the above hierarchy, it is also possible to compute $\epsilon_2=\frac{d\ln{\epsilon_1}}{d \mathcal{N}}=\frac{1}{\mathrm{H}}\big(\frac{\dot\epsilon_1}{\epsilon_1}\big),$ which implies $\epsilon_1\epsilon_2=d_{\mathrm{H}} \ddot{d_{\mathrm{H}}} =-\frac{1}{\mathrm{H}^2}\left(\frac{\ddot {\mathrm{H}}}{\mathrm{H}}-2\frac{\dot {\mathrm{H}}^2}{\mathrm{H}^2}\right)$. In the same manner higher slow-roll parameters may be computed. Equation (\ref{el}) essentially defines a flow in space with cosmic time being the evolution parameter, which is described by the equation of motion,

\be\label{el1}\epsilon_0\dot\epsilon_l-\frac{1}{d_{\mathrm{H}_i}}\epsilon_l\epsilon_{l+1}=0,~~~~l\geq 0.\ee
One can also check that (\ref{el1}) yields all the results obtained from the hierarchy defined in (\ref{el}), using the definition (\ref{dh}). Now, since we have an additional degrees of freedom appearing due to the coupling function $\alpha(\phi)$, we need to introduce yet another hierarchy of flow parameter \cite{SR},

\be\label{d1} \delta_1=4\dot \alpha \mathrm{H}\ll 1, ~~~~~\delta_{i+1}=\frac{d\ln|\delta_i|}{d\ln a}, ~~~~\text{with,} ~~~~i\geq 1.\ee
Clearly, for $i=1, ~\delta_2=\frac{d\ln|\delta_1|}{d \mathrm N}=\frac{1}{\delta_1}\frac{\dot\delta_1}{\dot{\mathrm N}},$ and $\delta_1\delta_2=\frac{4}{\mathrm{H}}\left(\ddot \alpha \mathrm{H}+\dot \alpha\dot {\mathrm{H}}\right),$ and so on. The slow-roll conditions therefore read $|\epsilon_i|\ll 1$ and $|\delta_i| \ll 1$, which are analogous to the standard slow-roll approximation.\\

We are now all set to check the viability of the action \eqref{AG} by comparing the inflationary parameters with currently released data sets \cite{Planck1, Planck2}. For this purpose, let us rearrange the ($^0_0$) and the $\phi$ variation equations of Einstein, viz., (\ref{M00}) and (\ref{Mphivariation}) respectively as,

\begin{center}
    \be\label{M1}\begin{split} & \alpha \mathrm{H}^2-\frac{\alpha\Lambda}{3} +\alpha'{\dot\phi}\mathrm{H}+96\gamma \mathrm{H}^8 \bigg{[}2\bigg{(}1+{1\over{ \mathrm{H}^2}}\big{(}\frac{\ddot {\mathrm{H}}}{\mathrm{H}}-2\frac{{\dot {\mathrm{H}}}^2}{\mathrm{H}^2}\big{)}\bigg{)}+7\bigg{(}1+\frac{\dot {\mathrm{H}}}{\mathrm{H}^2}\bigg{)}^2-8\bigg{(}1+\frac{\dot{\mathrm{H}}}{\mathrm{H}^2}\bigg{)}-2\bigg{]}\\&\hspace{3.0 in}-\frac{\dot\phi^2}{12} -\frac{V}{6}=0, \end{split} \ee
\end{center}
\be \label{M2} \begin{split} \ddot\phi +3\mathrm{H}\dot\phi=-V'-2\alpha'\Lambda+&6\alpha'\mathrm{H}^2\bigg{[}\bigg{(}1+\frac{\dot {\mathrm{H}}} {\mathrm{H}^2} \bigg{)}+1\bigg{]}.
\end{split}\ee
Since the coupling parameter is function of $\phi$ in general, one can express this as $\dot\alpha=\alpha'\dot\phi$, $\ddot\alpha=\alpha''\dot\phi^2+\alpha'\ddot\phi$. Due to the presence of coupling, one is required to apply some additional conditions as already mentioned, viz. $4|\dot\alpha|\mathrm{H}\ll 1$ and $|\ddot\alpha|\ll |\dot\alpha|\mathrm{H}$. In view of the slow-roll parameters  the above equations (\ref{M1}) and \eqref{M2} may therefore be expressed as,

\be \label{hir1}\begin{split}& \alpha \mathrm{H}^2-\frac{\alpha\Lambda}{3}-\frac{1}{4}\big{(}1+\delta_1\big{)}+\frac{1}{4}+96\gamma \mathrm{H}^8\bigg{[}2\big{(}1-\epsilon_1 \epsilon_2 \big{)}+7\big{(}1-\epsilon_1\big{)}^2-8\big{(}1-\epsilon_1\big{)}-2\bigg{]}-\bigg{(}\frac{\dot\phi^2}{12}+{V\over6}\bigg{)}=0,  \end{split}\ee
\be \label{phivarM} \begin{split} \ddot\phi +3\mathrm{H}\dot\phi=-V'-2\alpha'\Lambda+&6\alpha'\mathrm{H}^2\bigg{[}\big{(}1-\epsilon_1 \big{)}+1\bigg{]}.
\end{split}\ee
Under standard slow-roll conditions, the equations (\ref{hir1}) and (\ref{phivarM}) may be approximated to

\be\begin{split}\label{H2}96\gamma{\mathrm{H}^8}-\alpha \mathrm{H}^2+\frac{1}{6}\bigg{(}{V}+2\Lambda\alpha\bigg{)}+{{\dot\phi}^2\over 12}=0,\hspace{0.3in}
\ddot\phi+ 3\mathrm{H}\dot\phi =-V'-2\alpha'\Lambda+12\alpha'\mathrm{H}^2. \end{split}\ee
Substituting the form of the potential obtained in solution \eqref{param} and applying the Standard slow-roll conditions for coupled term, $\dot\phi^2\ll V$ and $|\ddot\phi|\ll 3\mathrm{H}\dot\phi$, we obtain,

\be\label{alphav} 6\alpha \mathrm{H}^2=-576\gamma{\mathrm{H}^8}-{2V_1\over \phi}-{V_0}~~~ \mathrm{and}~~~3\mathrm{H}\dot\phi =-2\mathrm{H}^2\phi.\ee
The above equations are still extremely difficult, if not impossible to handle due to the presence of ${\mathrm{H}}^8$ term. Let us therefore choose an additional relation,

\be \label{H8} {\mathrm{H}}^8=k^8 \phi^2.\ee
The above choice simply depicts that $\mathrm{H}$ varies extremely slowly during inflation. Shortly, we shall exhibit consistency of such a choice. Now, in view of the relation \eqref{H8}, equation \eqref{alphav} may be expressed as,

\be\label{alphav2} 6\alpha \mathrm{H}^2=-576\gamma k^8\phi^2-{2V_1\over \phi}-{V_0} \hspace{0.2in}\mathrm{and} \hspace{0.2in} 3\mathrm{H}\dot\phi =-2\mathrm{H}^2\phi.\ee
One can now calculate the slow-roll parameters from the above equations \eqref{alphav2} as:

\be\label{alphaveps}\begin{split}& \epsilon \equiv - {{\dot {\mathrm{H}}} \over {\mathrm{H}^2}} = \frac{(64\gamma{k^8}\phi^2-{V_1\over 9\phi})}{(96\gamma{k^8}\phi^2+{V_1\over 6\phi}+{{\mathrm{H}^2\phi^2}\over 12}+{V_0\over 6})}-\frac{\phi(192\gamma{k^8}\phi-{V_1\over 6\phi^2}+{\mathrm{H}^2\phi\over 6})}{3(96\gamma{k^8}\phi^2+{V_1\over 6\phi}+{V_0\over6}+{{\mathrm{H}^2\phi^2}\over 12})};\\& \eta\equiv \frac{\dot\epsilon}{{\mathrm{H}}\epsilon} = -{2\over3 }\phi \left({\epsilon'(\phi)\over \epsilon}\right);\end{split}\ee
\be\label{alphaN} \mathcal{N}(\phi)\simeq \int_{t_i}^{t_f}\mathrm{H}dt=\int_{\phi_i}^{\phi_f}\frac{\mathrm{H}}{\dot\phi}d\phi\simeq \int_{\phi_f}^{\phi_i}\frac{3(576\gamma{k^8}\phi^2+{2V_1\over \phi}+V_0)}{2\phi(576\gamma{k^8}\phi^2+{V_1\over \phi}+V_0+{{\mathrm{H}^2\phi^2}\over 2})}d\phi.\ee
In the following table-1, we have presented a set of data, varying the final value of the coupling parameter $\alpha_f$, and taking $\phi_i = 14~M_P$, which is the initial value of the scalar field, together with $k^8 = 1.22\times 10^{-11}~M_P^6$, $\gamma = -1.219\times 10^{8}~M_p^{-4}$. Not only that the inflationary parameters viz. scalar to tensor ratio ($r$) and the scalar tilt ($n_s$) show excellent fit with the currently released data set ($r < 0.055, ~ 0.96 < n_s < 0.97$) \cite{Planck1, Planck2}, but also the choice of the parameters fixes the energy scale of inflation $\mathrm{H}_* = 8.37\times10^{-2}M_P$ to the sub-Planckian scale. Note that, we get identical result of $\mathrm{H}_*$ both from equations \eqref{H8} and \eqref{alphaveps}, which proves consistency of the assumption \eqref{H8}.\\

\begin{figure}
\begin{center}
\begin{minipage}[h]{0.74\textwidth}
      \centering
\begin{tabular}{|c|c|c|c|c|c|}
\hline\hline
 ${\alpha_f}$ in ${M^2_p}$ & ${\phi_f}$~ in ${M_p}$ & $V_0$~in ${M^4_P}$ & $r$ & $n_s$ & $\mathrm N$ \\\hline
       8.90 & 8.89$\times10^{-11}$ &{-1.5} & 0.04335 & 0.9600 & 37 \\\hline
       8.32 & 9.52$\times10^{-11}$&{-1.4} & 0.04337& 0.9616 & 37 \\\hline
       7.72 & 1.03$\times10^{-10}$ &-1.3 & 0.04339 & 0.9631 & 36 \\\hline
       7.12 & 1.11$\times10^{-10}$ &-1.2 & 0.04342 & 0.9647 &36 \\\hline
       6.51 & 1.21$\times10^{-10}$ &-1.1 & 0.04345 & 0.9663 & 36 \\\hline
       5.90 & 1.33$\times10^{-10}$ &-1.0 &0.04347 & 0.9679 & 36 \\\hline
       5.34 & 1.48$\times10^{-10}$&-0.9 &0.04350 & 0.9694 & 36 \\\hline
       4.79 & 1.67$\times10^{-10}$ &-0.8 &0.04353 & 0.9709 & 36 \\\hline
       \hline
       \end{tabular}
       \captionof{table}{Under the choice $k^8=1.22\times10^{-11}M_P^6,~\gamma =-{1.219\times 10^8}M_P^{-4},~{V_1}=1\times 10^{-10}{M_P^5},~\phi_i=14.0 M_P;~\mathrm{so~that~}{\mathrm{H}^2}=0.0070{M_P^2}$, the inflationary parameters show excellent fit with currently available data.}
      \label{tab:table1}
       \end{minipage}%
   \end{center}
   \end{figure}

Since everything looks fine, let us now proceed to check whether scalar field executes oscillatory behaviour, required for graceful exit from inflation. We therefore consider the first equation of \eqref{H2} which is, $6\alpha \mathrm{H}^2=-576\gamma{\mathrm{H}^8}-{2V_1\over \phi}-{V_0}+{\dot\phi^2\over 2}$, where we have substituted the forms of $\alpha(\phi)$ and $V(\phi)$ from solution \eqref{param}. Now considering the Hubble parameter remains almost constant, one can replace it by the constant $\lambda$, without any loss of generality. Thus, we get,

\be \dot\phi^2=-\left(\lambda^2\phi^2-{2V_1\over\phi}\right).\ee\\
Integrating, we obtain,

\be \phi (t)=\frac{ \sqrt[3]{2V_1} \tan ^{\frac{2}{3}}\left(\frac{3}{2} \left(c_1 \lambda +\lambda  t\right)\right)}{\sqrt[3]{\lambda ^2 \{\tan ^2\left(\frac{3}{2} \left(c_1 \lambda +\lambda  t\right)\right)+1\}}},~~ \mathrm{or}~~\phi (t)=\frac{ \sqrt[3]{2V_1} \left\{-\tan \left(\frac{3}{2} \left(\lambda  t-c_1 \lambda \right)\right)\right\}{}^{2/3}}{\sqrt[3]{\lambda ^2 \{\tan ^2\left(\frac{3}{2} \left(\lambda  t-c_1 \lambda \right)\right)+1\}}}.\ee
Clearly,

\be \phi (t)={\sqrt[3]{{2V_1}\over \lambda^2}} \sin ^{\frac{2}{3}}\left(\frac{3}{2} \left(c_1 \lambda +\lambda  t\right)\right), ~~~~ \mathrm{or}
~~~\phi (t)=-{\sqrt[3]{{2V_1}\over \lambda^2}} \sin ^{\frac{2}{3}}\left(\frac{3}{2} \left(c_1 \lambda -\lambda  t\right)\right).\ee
Hence, the scalar field executes oscillatory behaviour ensuring graceful exit from inflationary regime.

\section{Matter dominated era:}

Since the action \eqref{AG} admits de-Sitter vacuum solution and shows excellent agreement with the recently released Planck's data, we therefore opt to study its behaviour, in the later matter-dominated epoch. It is well known that the oscillation of the scalar field at the end of inflation, creates  particles. These particles undergo collisions which reheats the universe and push it to the hot big bang phase. At this epoch, the universe is mostly dominated by radiation. At around redshift $z \approx 3200$, the matter-radiation equality is established, and thereafter at around $z \approx 1080$, radiation is decoupled from matter, which is the onset of a pressure-less dust dominated era. These epochs are best described by Friedmann solutions $a \propto \sqrt t$ in the radiation dominated era, and $a \propto t^{2\over 3}$ in the matter dominated era. However, very recently at around $z \approx 1$, the universe starts accelerating yet again. It is therefore suggestive to explore the behaviour of the present action \eqref{AG} in these epochs. Unfortunately, setting $\beta = 0$, the field equations \eqref{Mzvariation}, \eqref{M00}, \eqref{Mphivariation}, being expressed in view of the forms of the $\alpha(\phi)$ and $V(\phi)$ \eqref{param}, and using the Bianchi identity,

\be\begin{split} \label{fe1} & 2\left(\alpha_0-{\alpha_1\over \phi}-{\phi^2\over 12}\right)\left({\ddot z\over z}-{\dot z^2\over 4z^2} - \Lambda\right)+2\left({\alpha_1\over \phi^2}-{\phi\over 6}\right)\left({\ddot\phi}+{\dot\phi}{\dot z\over z} \right)-2\left({2\alpha_1\over \phi^3}+{1\over 6}\right){\dot\phi^2}\\&\hspace{0.3 in}+12\gamma\left[{\dot z^4{\ddddot z}\over z^5}+{8\dot z^3\ddot z\dddot z\over z^5}-{9\dot z^5\dddot z\over z^6}+{6\dot z^2\ddot z^3\over z^5}-{135 \dot z^4\ddot z^2 \over 4 z^6}+{159 \dot z^6\ddot z\over 4z^7} - {195 \dot z^8\over 16 z^8}\right] \\&\hspace{0.3 in} = -\omega{\rho_0}{a^{-3(1+\omega)}} -{1\over 2}\dot\phi^2 +\left[-1152\gamma\lambda^8-2{V_1\over \phi}-V_0+2\Lambda\left(96\gamma\lambda^6+\frac{V_0}{6\lambda^2}+\frac{V_1}{6\lambda^2\phi}+ {\phi^2\over 12}\right)\right];\\&
2\left(\alpha_0-{\alpha_1\over \phi}-{\phi^2\over 12}\right)\left({3\dot z^2\over 4z^2}-\Lambda\right)+{3\dot\phi \dot z\over z}\left({\alpha_1\over \phi^2}-{\phi\over 6}\right)+ 18\gamma\left[{\dot z^5 \dddot z\over z^6} + {3\dot z^4 \ddot z^2\over 2z^6} - {9 \dot z^6\ddot z\over 2 z^7} + {15 \dot z^8\over 8 z^8}\right]\\& = {\rho_0}{a^{-3(1+\omega)}} + {1\over 2}\dot\phi^2 +\left[-1152\gamma\lambda^8-2{V_1\over \phi}-V_0+2\Lambda\left(96\gamma\lambda^6+\frac{V_0}{6\lambda^2}+\frac{V_1}{6\lambda^2\phi}+ {\phi^2\over 12}\right)\right] ;\\&
\ddot\phi + {3\over 2}{\dot z\over z}\dot\phi-{3\ddot z\over z }\left({\alpha_1\over \phi^2}-{\phi\over 6}\right)+2\Lambda\left({\alpha_1\over \phi^2}-{\phi\over 6}\right) +\left(2-{\Lambda\over 3\lambda^2}\right){V_1\over \phi^2}+{{\Lambda\phi}\over 3} = 0.\end{split}\ee
\be \label{Bianchi} \dot \rho + 3H(\rho+p) = 0,\ee
do not give analytical solution in any form. On the contrary, considering $\alpha =$ constant, it is possible to explore analytical solution in the form $a \propto \sqrt t$ in the radiation dominated era ($p = {1\over 3}\rho$), and $a \propto t^{2\over 3}$ in the pressure-less dust ($p = 0$) era, while the scalar field $\phi \propto t^{-3}$ in both the cases, provided the potential takes the form $V = -2\alpha\Lambda - V_1 \phi^{8\over 3}$. The Friedmann-like behaviour of such a complicated action is definitely encouraging, but, this does not work, since we have already mentioned that for constant $\alpha$, de-Sitter vacuum solution is not realized. We therefore study the behaviour of the field equations under following theoretical inspection.\\

Firstly, let us mention that different modified theories of gravity have been suggested as an alternative to the dark energy comprising a scalar field, since such a field does not exist in the present universe. Here, we note that the scalar field at the end of inflation has come down to a reasonably small value ($\phi_f \sim 10^{-11} M_P$). If we therefore assume that the rest of it has been mostly expended in creating baryons, and left out had been red-shifted very fast, then we can get rid of the scalar field, and consider both $\alpha$ and $V = -U_0$ become constant. In that case, the above set of field equations take the form:

\be\begin{split} \label{fe1} & 2\alpha\left({\ddot z\over z}-{\dot z^2\over 4z^2}\right) +12\gamma\left[{\dot z^4{\ddddot z}\over z^5}+{8\dot z^3\ddot z\dddot z\over z^5}-{9\dot z^5\dddot z\over z^6}+{6\dot z^2\ddot z^3\over z^5}-{135 \dot z^4\ddot z^2 \over 4 z^6}+{159 \dot z^6\ddot z\over 4z^7} - {195 \dot z^8\over 16 z^8}\right] = - p +(2\alpha\Lambda - U_0),\\&
2\alpha\left({3\dot z^2\over 4z^2}\right) + 18\gamma_0\left[{\dot z^5 \dddot z\over z^6} + {3\dot z^4 \ddot z^2\over 2z^6} - {9 \dot z^6\ddot z\over 2 z^7} + {15 \dot z^8\over 8 z^8}\right] = \rho + (2\alpha\Lambda - U_0).\end{split}\ee
Now, coefficient of $\gamma$ comes from the contribution of $\mathcal{G}^2$ term, which is negligible at the later stage of cosmic evolution. For example, if we seek solution in the form $a \propto t^n$, then this term falls as $t^{-8}$, while the first term evolve as $t^{-2}$. Thus, we can neglect contribution of $\mathcal{G}^2$ term as time past a little more than a second. In that case, the above equations are simply the Friedmann equations, provided $U_0 = 2\alpha \Lambda$, and cosmic evolution follows according to the standard model of cosmology. To handle the dark energy issue, one has to incorporate an  additional term in the action in the form $\mathcal{G}^m$, where $m < {1\over 2}$, as mentioned in the introduction. Since such a term contributes negligibly in the early universe, we disregarded it for simplicity. However, at the late stage of cosmic evolution, such a term dominates over others, and late time acceleration is realized. Crossing of phantom divide line is also possible for $m < 0$, as expatiated in a host of articles already cited in the introduction.

\section{Conclusion:}

A host of modified theories of gravity has been proposed in recent times as alternatives to the dark energy issue. It is therefore suggestive to check viabilities of different higher order modified theories of gravity in the context of very early universe. The claim regarding the prevailing $F(\mathcal{G})$ theory is that, $F(\mathcal{G}) = \alpha \mathcal{G}^n + \beta \mathcal{G}^m$ theory of gravity can unify early inflation with late time acceleration, provided $n > 1$ required for inflation, and $m < {1\over 2}$ for late-time cosmic acceleration. Here, we therefore opted to test inflation in view of action \eqref{A}. Effectively $\mathcal{G}^2$ term behaves like $R^4$ term, at least in the background of isotropic and homogeneous cosmological model under consideration. We therefore have included $R^2$ term, with the idea that a reasonable action must contain different orders systematically. We have also disregarded lower order term in $\mathcal{G}$, to avoid unnecessary complication, since such a term remains subdominant in the early universe. Although, inflation is a quantum theory of perturbation, it may be studied in view of the classical field equations, provided the theory admits a smooth crossover from quantum to the classical domain, under some appropriate semiclassical approximation. Further, it is often stated that in the absence of a complete quantum theory of gravity, quantum cosmology may be probed to unveil certain physical insights in the Planck's era. Due to diffeomorphic invariance, the gravitational Hamiltonian is constrained to vanish. Thus the concept of time ceases in the quantum domain, which is a major setback of General Theory of Relativity (GTR). This led Hartle–Hawking \cite{HH}, Hawking and Page \cite{HP} and also Vilenkin \cite{V1, V2}, to put forward different proposals to interpret the wave function of the universe associated with GTR. The beauty of incorporating higher-order terms in the gravitational action is: an internal parameter plays the role of time, and as a result, standard quantum mechanical probability interpretation is envisaged. \\

As mentioned, in order to study inflation, it is necessary to check if the semiclassical wavefunction is oscillatory about a classical de-Sitter solution. For this purpose, canonically quantization is primarily required, canonical formulation being its precursor. We have followed it systematically in the present work. There are some important outcomes. Firstly, the present $F(R, \mathcal{G})$ gravity theory admits a classical de-Sitter vacuum solution, provided $\alpha = \alpha(\phi)$. Next, curvature invariant terms of different higher orders are not allowed in the gravitational action, at least from the point of view of a viable quantum theory (hermitian effective Hamiltonian operator and existence of continuity equation). We believe that this physical insight being revealed from quantum cosmology, might lead to a new understanding towards formulating a reasonable theory of quantum gravity. Finally, only if the scalar field ceases to evolve during matter dominated era, Friedmann solutions ($a \propto \sqrt t$ in radiation dominated era and $a \propto t^{2\over 3}$ in the pressureless dust era) are realized. A lower power of the Gauss-Bonnet term $\mathcal{G}$, might thereafter lead to accelerated expansion of the universe at the late stage.

\appendix
\section{Semiclassical approximation:}

In the appendix, we briefly compute semiclassical wavefunction of the quantum equation (\ref{Mqn}. Let us express equation (\ref{Mqn}) as,
\begin{center}
\be \label{qh1cs4} \begin{split} &-\frac{z^{\frac{9}{2}}\hbar^2}{36\gamma x^5}\bigg{(}\frac{\partial^2}{\partial x^2} +\frac{n}{x}\frac{\partial}{\partial x}\bigg{)}\Psi-\frac{\hbar^2}{2xz^{\frac{3}{2}}}\frac{\partial^2\Psi}{\partial \phi^2}-{i\hbar}\frac{\partial\Psi}{\partial z}+\frac{3i\hbar } {z}\bigg{(}-\frac{\alpha_1}{\phi^2}+{\phi\over 6 }\bigg{)} \frac{\partial\Psi}{\partial \phi}\\&\hspace{3.0 in} +\frac{3i\hbar}{z}\bigg{(}\frac{\alpha_1}{\phi^3}+{1\over 12}\bigg{)}\Psi +\mathcal{V}\Psi=0, \end{split}\ee
\end{center}
where,
\be\begin{split}& \mathcal{V}=\left[\frac{9x}{2\sqrt z} \bigg{(} \frac{\alpha_1^2}{\phi^4}-\frac{\alpha_1}{3\phi}+{\phi^2\over 36 }\bigg{)} -\frac{3 x}{2\sqrt z}\bigg{(}\alpha_0-\frac{\alpha_1}{\phi}-{\phi^2\over 12}\bigg{)}+\frac{135\gamma x^7}{28z^{13\over 2}} +\frac{z^{3\over 2}}{x}\bigg{(}-1152\gamma\lambda^8-2{V_2\over \phi}-V_1\bigg{)}\right]\end{split} \ee
Equation (\ref{qh1cs4}) may be treated as time independent Schr{\"o}dinger equation with three variables ($x$, $z$, $\phi$), and therefore as usual, let us seek the solution of equation (\ref{qh1cs4}) in the standard form,

\be\label{Psi4}\Psi = \Psi_0(x,z,\phi)e^{\frac{i}{\hbar}S(x,z,\phi)},\ee
where, $\Psi_0$ is a slowly varying function, and expand $S$ in power series of $\hbar$ as,
\be\label{S4} S = S_0(x,z,\phi) + \hbar S_1(x,z,\phi) + \hbar^2S_2(x,z,\phi) + .... .\ee\
As a result one can compute,
\begin{center}

    \be \label{psiall} \begin{split} & \Psi_{,x}=\Psi_{0,x}e^{{i\over \hbar}S}+{i\over \hbar}\bigg[S_{0,x}+\hbar S_{1,x}+\hbar^2 S_{2,x}+\mathcal{O}(\hbar)  \bigg]\Psi_0 e^{{i\over \hbar}S};\\& \Psi_{,xx}=2{i\over \hbar}\bigg[S_{0,x}+\hbar S_{1,x}+\hbar^2 S_{2,x}+\mathcal{O}(\hbar)  \bigg]\Psi_{0,x} e^{{i\over \hbar}S}+{i\over \hbar}\bigg[S_{0,xx}+\hbar S_{1,xx}+\hbar^2 S_{2,xx}+\mathcal{O}(\hbar)  \bigg]\Psi_0 e^{{i\over \hbar}S}\\&+\Psi_{0,xx}e^{{i\over \hbar}S}- {1\over \hbar^2}\bigg[S_{0,x}^2+\hbar^2 S_{1,x}^2+\hbar^4 S_{2,x}^4+2\hbar S_{0,x} S_{1,x}+2\hbar^2 S_{0,x} S_{2,x}+2\hbar^3 S_{1,x}S_{2,x}+\mathcal{O}(\hbar)\bigg]\Psi_0 e^{{i\over \hbar}S} ;\\& \Psi_{,\phi}=\Psi_{0,\phi}e^{{i\over \hbar}S}+{i\over \hbar}\bigg[S_{0,\phi}+\hbar S_{1,\phi}+\hbar^2 S_{2,\phi}+\mathcal{O}(\hbar)  \bigg]\Psi_0 e^{{i\over \hbar}S};\\& \Psi_{,\phi\phi}=2{i\over \hbar}\bigg[S_{0,\phi}+\hbar S_{1,\phi}+\hbar^2 S_{2,\phi}+\mathcal{O}(\hbar)  \bigg]\Psi_{0,\phi} e^{{i\over \hbar}S}+{i\over \hbar}\bigg[S_{0,\phi\phi}+\hbar S_{1,\phi\phi}+\hbar^2 S_{2,\phi\phi}+\mathcal{O}(\hbar)  \bigg]\Psi_0 e^{{i\over \hbar}S}\\&+\Psi_{0,\phi\phi}e^{{i\over \hbar}S}- {1\over \hbar^2}\bigg[S_{0,\phi}^2+\hbar^2 S_{1,\phi}^2+\hbar^4 S_{2,\phi}^4+2\hbar S_{0,\phi} S_{1,\phi}+2\hbar^2 S_{0,\phi} S_{2,\phi}+2\hbar^3 S_{1,\phi}S_{2,\phi}+\mathcal{O}(\hbar)\bigg]\Psi_0 e^{{i\over \hbar}S} ;\\& \Psi_{,z}=\Psi_{0,z}e^{{i\over \hbar}S}+{i\over \hbar}\bigg[S_{0,z}+\hbar S_{1,z}+\hbar^2 S_{2,z}+\mathcal{O}(\hbar)  \bigg]\Psi_0 e^{{i\over \hbar}S}. \end{split}\ee
\end{center}
In the above, ‘comma’ everywhere in the suffix represents derivative. Now inserting the expressions (\ref{S4}) and (\ref{Psi4}) after taking appropriate derivatives in equation (\ref{qh1cs4}) and equating the coefficients of different powers of $\hbar$ to zero, one obtains the following set of equations (upto second order)
\be\begin{split}
&\frac{z^\frac{9}{2}}{36\gamma x^5}S_{0,x}^2 + \frac{S_{0,\phi}^2}{2xz^{\frac{3}{2}}} + S_{0,z}+\frac{3} {z}\bigg{(}-\frac{\alpha_1}{\phi^2}+{\phi\over 6 }\bigg{)} S_{0,\phi} + \mathcal{V}(x,z,\phi) = 0, \label{S0}\end{split}\ee

\be\begin{split}
&-\frac{z^\frac{9}{2}}{36\gamma x^5}\bigg[ \bigg(iS_{0,\phi\phi}-2S_{0,x}S_{1,x}+{in\over x}S_{0,x}\bigg)\Psi_0+2iS_{0,x} \Psi_{0,x}\bigg]-{1\over {2xz^{\frac{3}{2}}}}\bigg[ \bigg(iS_{0,\phi\phi}-2S_{0,\phi}S_{1,\phi}\bigg)\Psi_0\\&+2iS_{0,\phi} \Psi_{0,\phi}\bigg]+S_{1,z}\Psi_0-i\Psi_{0,z}+{3i\over z}\bigg(-{\alpha_1\over \phi^2}+{\phi\over 6}\bigg)\bigg( S_{1,\phi}\Psi_0+\Psi_{0,\phi}\bigg)+{3i\over z}\bigg( {\alpha_1\over \phi^3}+{1\over 12}\bigg)\Psi_0
= 0, \label{S1}\end{split}\ee
\be\begin{split}
&-\frac{z^\frac{9}{2}}{36\gamma x^5}\bigg[\bigg(iS_{1,xx}-S_{1,x}^2-2S_{0,x}S_{2,x}+{in\over x}S_{1,x}\bigg)\Psi_0+\Psi_{0,xx}+2iS_{1,x}\Psi_{0,x}+{n\over x}\Psi_{0,x}\bigg]\\& -{1\over 2xz^{3\over 2}}\bigg[\bigg( iS_{1,\phi\phi}-S_{1,\phi}^2-2S_{0,\phi}S_{2,\phi}\bigg)\Psi_0+\Psi_{0,\phi\phi}+2iS_{1,\phi}\Psi_{0,\phi}\bigg]+S_{2,z}\Psi_{0}- {3\over z}\bigg(-{\alpha_1\over \phi^2}+{\phi\over 6} \bigg)S_{2,\phi}\Phi_0=0.
\label{S2}\end{split}\ee
These equations \eqref{S0}-\eqref{S2} are to be solved successively to find $S_0$, $S_1$ and $S_2$ and so on. Now identifying $S_{0,x}$, as $p_x$, $S_{0,\phi}$ as $p_{\phi}$ and $S_{0,z}$ as $p_z$; the classical Hamiltonian constraint equation $\mathcal{H}=0$, presented in equation \eqref{woR2} for $k=0$, may be recovered from equation \eqref{S0}, which is therefore identified as the Hamilton-Jacobi equation.
The Hamilton-Jacobi function, $S_0(x,\phi,z)$ therefore is expressed as,
\be \label{S0S} S_0=\int {p_x}dx +\int {p_{\phi}}d\phi+\int p_z dz,\ee
apart from a constant of integration which may be absorbed in $\Psi_0.$ It is possible to evaluate the integrals
in the above expression, using the classical solutions \eqref{aphi}, \eqref{param} together with the definitions of momenta presented in \eqref{p1} and \eqref{pz}, keeping the relation $\dot z=Nx.$ For the present purpose, we fix the gauge $N=1.$
Thus, the expressions of momenta are found as
\be p_x=576\sqrt{2}\gamma \lambda^{11\over 2}\sqrt{x};~~p_{\phi}=-{6\alpha_1 \lambda a_0^3\phi_0^3\over \phi^5};~~p_z=-6\alpha_0 \lambda \sqrt{z}+{9\alpha_1\lambda\over a_0\phi_0}z-{10368\gamma \lambda^7\over 7}\sqrt{z}. \ee
The explicit form of $S_0$ may then be computed to find,

\be S_0=-4\alpha_0\lambda z^{3\over 2}+{6\alpha_1 \lambda z^2\over a_0\phi_0}+{3840\gamma \lambda^7 z^{3\over 2}\over 7}.\ee
Therefore, the semiclassical wavefunction is found as,
\be\label{MPsi4}\Psi = \Psi_0e^{\frac{i}{\hbar}S(x,z,\phi)}=\Psi_{01}e^{{i\over \hbar}\Big[-4\alpha_0\lambda z^{3\over 2}+{6\alpha_1 \lambda z^2\over a_0\phi_0}+{3840\gamma \lambda^7 z^{3\over 2}\over 7}\Big]},\ee
where, \be \Psi_{01}=\Psi_0e^{-F(z)}.\ee

\end{document}